\documentclass[a4paper]{article}
\title{Bosonic sector of $D=11$ superstring action and the
critical dimension. Toy model.}
\author{A.A. Deriglazov\thanks{alexei@fisica.ufjf.br ~ On leave of
absence from Dept. Math. Phys., Tomsk Polytechnical University,
Tomsk, Russia}}
\date{Instituto de F\'\i sica, Universidade Federal do Rio de Janeiro,\\
Rio de Janeiro, Brasil.}
\begin{document}
\maketitle
\large
\begin{abstract}
Bosonic model inspired by $D=11$ superstring action is 
investigated. An appropriate set of variables is find, in which 
the light-cone quantization turns out to be possible. 
It is shown that anomaly terms in the
algebra of the light-cone Poincare generators are absent for the
case $D=27$.
\end{abstract}

\noindent

Construction of $D=11$ Green-Schwarz type superstring action presents
a nontrivial problem already at the classical level. The reason is that
only for the dimensions $D=3,4,6,10$ the action is invariant under the
local $\kappa$-symmetry (as well as under the global supersymmetry) [1].
Recently it was recognized [2-6] that the problem can be resolved if one
introduces an additional vector variable $n^N$ into the formulation.
The corresponding $D=11$ action (which incorporates $n^N(\tau, \sigma)$
as the dynamical variable) was suggested in [3]. Similarly to the
Green-Schwarz construction, the action has $\kappa$-symmetry which allows
one to remove half of fermionic coordinates and supply free dynamics for
the physical variables as well as the discrete mass spectrum [3,4]. 
Moreover, $n^N$-independent part of spectrum
(being classified with respect to $SO(1,9)$ group) was identified with
the type IIA superstring states. For the massless level classified with 
respect to $SO(1,10)$ group one gets the supergravity multiplet in 
$D=11$ [7-9]. The other states (presented on each
mass level) may correspond to the states of the uncompactified M-theory
limit [9,10]. Due to these properties one hopes that such a kind theory
can be reasonable extension of the Green-Schwarz action to the
case $D=11$.

The aim of this work is to investigate some quantum properties of the
theory. It will be demonstrated that light-cone quantization of the
bosonic sector is possible (the corresponding Lagrangian will be 
discussed below), which allows one to compute algebra of the
light-cone Poincare generators. We show that anomaly terms in the
algebra are absent for the case $D=27$. Fermionic sector of $D=11$
superstring action consist of $D=11$ Majorana spinor which can be
decomposed on a pair of the Majorana - Weyl spinors of an opposite
chirality with respect of $SO(1,9)$ group. From this fact and from
the result $D=27$ for the bosonic sector one expects that the
critical dimension for the superstring presented in [3,4] is $D=11$.

Let $\pi^N$ is zero mode of $n^N(\tau, \sigma)$ [5], and the corresponding
canonically conjugated variable will be denoted as $\tilde y^N$. Our
starting point is the Virasoro constraints
\begin{eqnarray}\label{1}
L_n=\frac 12\sum_{\forall k}\tilde\alpha^N_{n-k}\tilde\alpha^N_k=0, \quad 
\bar L_n=\frac 12\sum_{\forall k}\tilde{\bar\alpha}^N_{n-k}
\tilde{\bar\alpha}^N_k=0,
\end{eqnarray}
accompanied by the first class constraint
$\pi^N\pi^N=-\varepsilon=const$ and by the following second class system
\begin{eqnarray}\label{2}
\pi^N\tilde\alpha^N_n=0, \qquad \pi^N\tilde{\bar\alpha}^N_n=0, \qquad
n \ne 0;
\end{eqnarray}
\begin{eqnarray}\label{3}
\pi^N\tilde\alpha_0^N=0, \qquad \pi^N\tilde x^N=0. 
\end{eqnarray}
Below we will omit expressions for the left moving oscillators
$\tilde{\bar\alpha}^N$. The cases of $SO(1,D-1)$ and $SO(2,D-2)$ group
will be considered simultaneously:
$\eta^{NM}=(\eta^{\mu\nu}, \eta^{D-1,D-1}\equiv\eta), ~ \eta=\pm 1, ~ 
\eta^{\mu\nu}=(-,+,\ldots ,+), ~ \mu, \nu=0,1, \ldots ,D-2$. 
The parameters $\varepsilon, \eta$ are not fixed (except the restrictions 
which follows from the constraints) throughout the work, but is 
expected to be fixed in the supersymmetric version [4]. The string
tension is chosen to be $T=\frac{1}{4\pi}$ such that
$\tilde\alpha_0^N=-\tilde{\bar\alpha}_0^N=\tilde p^N$. The system (\ref{1}),
(\ref{2}) can be obtained by means of partial fixation of gauge for
the bosonic constraints presented in the theory [6]. As it was shown
in [3,4], these constraints (and the corresponding terms in the action)
are essential for establishing of the $\kappa$-symmetry. Below we
present also an action which leads to the complete system
(\ref{1})-(\ref{3}).

$D$-dimensional Poincare generators are realized as
\begin{eqnarray}\label{4}
{\bf P}^N=-\tilde p^N, \qquad 
{\bf J}^{MN}=
\tilde x^{\left[ M\right.} \tilde p^{\left. N\right]}
+iS^{MN}+i\bar S^{MN}+\tilde y^{\left[ M\right.}\pi^{\left. N\right]}, \cr
S^{MN}\equiv \sum_{n=1}^{\infty}\frac 1n
\tilde\alpha^{\left[ M\right.}_{-n}\tilde\alpha^{\left. N\right]}_n.
\end{eqnarray} 
From Eq.(\ref{2}) it follows that one component of each oscillator is
gauge degree of freedom. So one expects that only the remaining
$D-1$ components will give contribution into the anomaly terms, such
that the condition of absence of the anomaly will be: $D-1=26$. We
support this suggestion by direct calculations.

To quantize the theory we follow to the standard prescription [11,12].
The second class constraints (\ref{2}), (\ref{3}) can be taken into
account by means of introduction of the corresponding Dirac bracket.
The non zero brackets for our basic variables turn out to be
\begin{eqnarray}\label{5}
\left\{\tilde x^N, \tilde p^M\right\}=\Pi^{NM}\equiv\eta^{NM}-
\frac{1}{\pi^2}\pi^N\pi^M,\nonumber\\
\left\{\tilde\alpha^N_n, \tilde\alpha^M_k\right\}=
in\delta_{n+k,0}\Pi^{NM},  \nonumber\\
\left\{\tilde y^N, \pi^M\right\}=\eta^{NM}, \nonumber\\
\left\{ \tilde y^N, \tilde y^M\right\}=
-\frac{1}{\pi^2}\tilde x^{\left[ N\right.}\tilde p^{\left. M\right]}-
i\sum_{n=1}^\infty\frac{1}{n\pi^2}
(\tilde\alpha^{\left[ N\right.}_{-n}\tilde\alpha^{\left. M\right]}_n+
\tilde{\bar\alpha}^{\left[ N\right.}_{-n}
\tilde{\bar\alpha}^{\left. M\right]}_n), \\
\left\{ \tilde x^M, \tilde y^N\right\}=
\frac{1}{\pi^2}\pi^M\tilde x^N, \qquad
\left\{ \tilde p^M, \tilde y^N\right\}=
\frac{1}{\pi^2}\pi^M\tilde p^N,  \nonumber\\
\left\{ \tilde\alpha^M_n, \tilde y^N\right\}=
\frac{1}{\pi^2}\pi^M\tilde\alpha^N_n\nonumber,
\end{eqnarray}
and the same expressions for the left moving oscillators
$\tilde{\bar\alpha}^N_n$. Now Eqs.(\ref{2}),(\ref{3}) can be solved
\begin{eqnarray}\label{6}
\tilde z^{D-1}=-\frac{\eta}{\pi^{D-1}}\pi^\nu\tilde z^\nu,
\end{eqnarray}
where $\tilde z=(\tilde x, \tilde p, \tilde\alpha_n,
\tilde{\bar\alpha}_n)$. Since brackets for the remaining variables
$\tilde x^\nu, \tilde p^\nu, \tilde\alpha^\nu_n, \tilde y^N, 
\pi^N$ are rather complicated,
it is convenient to simplify them by means of an appropriate variable
change. The change turns out to be
\begin{eqnarray}\label{7}
x^\mu=\tilde x^\mu+c\pi^\mu(\pi \tilde x), \qquad
p^\mu=\tilde p^\mu+c\pi^\mu(\pi\tilde p), \cr
\alpha^\mu_n=\tilde\alpha^\mu_n+c\pi^\mu(\pi\tilde\alpha_n), \cr
y^\mu=\tilde y^\mu+c\left[(\pi\tilde x)\tilde p^\mu-
(\pi\tilde p)\tilde x^\mu\right]+ \cr
ic\sum_{n=1}^\infty\left[\frac{1}{n}
(\pi\tilde\alpha_{-n})\tilde\alpha^\mu_n+
(n\leftrightarrow -n)\right]+(\tilde{\bar\alpha}- sector), \cr
y^{D-1}\equiv\tilde y^{D-1},
\end{eqnarray}
where from now $(\pi\tilde x)\equiv\pi^\mu\tilde x^\mu$, and so on.
The factor $c$ is any solution of the equation
$(\pi^2)c^2+2c-\eta(\pi^{D-1})^{-2}=0$, thus
\begin{eqnarray}\label{8}
c=\frac{1}{\pi^2}\left[-1\pm\frac{(\eta\pi^N\pi^N)^
{\frac{1}{2}}}{\pi^{D-1}}\right].
\end{eqnarray}
The new variables obey to the canonical brackets
\begin{eqnarray}\label{9}
\{x^\mu, p^\nu\}=\eta^{\mu\nu}, \quad \{y^N, \pi^M\}=\eta^{NM}, \quad 
\{\alpha^\mu_n, \alpha^\nu_k\}=
in\eta^{\mu\nu}\delta_{n+k,0}.
\end{eqnarray}
Eq.(\ref{7}) is invertible, an opposite change has the same form and
can be obtained from Eq.(\ref{7}) by means of substitution
$z\leftrightarrow\tilde z, ~ y\leftrightarrow\tilde y, ~ 
c\mapsto \bar c$, where
\begin{eqnarray}\label{10}
\bar c=\frac{1}{\pi^2}\left[-1\pm{\pi^{D-1}}(\eta\pi^N\pi^N)^
{-\frac{1}{2}}\right].
\end{eqnarray}
Note that a variable change which leads to Eq.(\ref{9}) is not unique.
For example (for the Dirac bracket which corresponds to Eq.(\ref{2}))
the following simple change
\begin{eqnarray}\label{11}
\alpha^\mu_n=\tilde\alpha^\mu_n-\pi^\mu\frac{\tilde\alpha^{D-1}_n}
{\pi^{D-1}}, \qquad \alpha^\mu_{-n}\equiv\tilde\alpha^\mu_{-n}, \cr
y^N=\tilde y^N+
i\sum_{n=1}^\infty\frac{1}{n\pi^{D-1}}
(\tilde\alpha^N_{-n}\tilde\alpha^{D-1}_n+
\tilde{\bar\alpha}^N_{-n}\tilde{\bar\alpha}^{D-1}_n),
\end{eqnarray}
gives also the canonical brackets for the new variables. The problem is
that the Virasoro constraints, being rewritten in terms of these
variables, will contain products of $\alpha^-_n$ oscillators:
$L_n\sim p^+\alpha^-_n+\frac 12(\pi^+)^2\sum_{k=0}^{n-1}
\alpha^-_{n-k}\alpha^-_k+\ldots$. It do not allows one to resolve the
constraints in the light-cone gauge. In contrast, our change
(\ref{7}) leads to the "linearised" form of the constraints. Namely,
substitution of Eqs.(\ref{6}), (\ref{7}) into Eq.(\ref{1}) gives the
expressions
\begin{eqnarray}\label{12}
L_n=\frac 12\sum_{\forall k}\alpha^\mu_{n-k}\alpha^\mu_k=0, \quad 
\bar L_n=\frac 12\sum_{\forall k}\bar\alpha^\mu_{n-k}
\bar\alpha^\mu_k, \cr
L_0+\bar L_0=(p^\mu)^2+\sum_{k=1}^\infty(\alpha^\mu_{-k}\alpha^\mu_k+
\bar\alpha^\mu_{-k}\bar\alpha^\mu_k)=0,
\end{eqnarray}
\begin{eqnarray}\label{13}
L_0-\bar L_0=\sum_{k=1}^\infty(\alpha^\mu_{-k}\alpha^\mu_k-
\bar\alpha^\mu_{-k}\bar\alpha^\mu_k)=0, \qquad \mu=0,1,\ldots ,D-2
\end{eqnarray}
which contain the variables $p^\mu, \alpha^\mu_n, \bar\alpha^\mu_n$
only. Now the light-cone quantization can be carried out in the
standard form [7,13,14]. One imposes the gauge
$x^+=\alpha^+_n=\bar\alpha^+_n=0$, then the
variables $p^-, \alpha^-_n, \bar\alpha^-_n$ can be expressed through
the remaining
(D-3)-dimensional oscillators $\alpha^i_n, ~ \bar\alpha^i_n, ~ 
i=1,2,\ldots ,D-3$
\begin{eqnarray}\label{14}
p^-=\frac{1}{2p^+}(L_0^{tr}+\bar L_0^{tr}-a), \quad 
\alpha^-_n=\frac{1}{p^+}L_n^{tr}, \quad
\bar\alpha^-_n=-\frac{1}{p^+}\bar L_n^{tr}, \cr
L_n^{tr}=\frac 12\sum_{\forall k} \alpha^i_{n-k}\alpha^i_k, \qquad
L_0^{tr}=\frac 12(p^i)^2+\sum_{k=1}^{\infty}\alpha^i_{-k}\alpha^i_k.
\end{eqnarray}
The oscillators are arranged in the normal order, the corresponding
normal ordering constant $a$ is included into the expression for $p^-$.

By using of Eqs.(\ref{4}), (\ref{7}), (\ref{14}) one obtains the 
light-cone Poincare generators which can be presented as
\begin{eqnarray}\label{15}
{\bf P}^\mu={\bf P}^\mu_{(D-1)}+\bar c\pi^\mu(\pi{\bf P}_{(D-1)}), 
\nonumber\\
{\bf J}^{\mu\nu}={\bf J}^{\mu\nu}_{(D-1)}+y^{\left[\mu\right.}
\pi^{\left. \nu\right]}, \nonumber\\
{\bf P}^{D-1}=\pm\eta (\eta\pi^N\pi^N)^{-\frac 12}
(\pi{\bf P}_{(D-1)}), \\
{\bf J}^{\mu D-1}=c\pi^{D-1}\pi^\nu{\bf J}^{\nu\mu}_{(D-1)}+
y^{\left[\mu\right.}\pi^{\left. D-1\right]}.\nonumber
\end{eqnarray}
The quantities ${\bf P}_ {(D-1)}, {\bf J}_{(D-1)}$ coincide with the
standard $(D-1)$-dimensional Poincare generators of the closed string
\begin{eqnarray}\label{16}
{\bf P}^\mu_{(D-1)}=-p^\mu, \qquad 
{\bf J}^{\mu\nu}_{(D-1)}=
x^{\left[\mu\right.}p^{\left. \nu\right]}
+iS^{\mu\nu}+i\bar S^{\mu\nu}, \cr
S^{\mu\nu}=\sum_{n=1}^{\infty}\frac 1n
\alpha^{\left[ \mu\right.}_{-n}\alpha^{\left. \nu\right]}_n,
\end{eqnarray} 
where it is implied that Eq.(\ref{14}) was substituted. Note that
$-M^2=({\bf P}^\mu)^2+\eta({\bf P}^{D-1})^2\equiv (p^\mu)^2$ from which
it follows that the last from Eq.(\ref{12}) actually gives
the mass formula. Thus, in terms of the new variables (\ref{7}),
$D$-dimensional Poincare generators of the theory is presented through
the usual $(D-1)$-dimensional one. It makes analysis of the anomaly
terms an easy task. By construction, commutators of the quantities
(\ref{15}) form $D$-dimensional Poincare algebra modulo to the terms
which can arise in the process of reordering of oscillators to the
normal form. The quantities (\ref{15}) have the following structure:
$A(y, \pi)+B(\pi)C_{(D-1)}(x,p,\alpha,\bar\alpha)$, where $C_{(D-1)}$
represents the generators (\ref{16}). Then structure of any commutator
is
\begin{eqnarray}\label{17}
\left[A^1(y,\pi), A^2(y,\pi)\right]+\left[A(y,\pi), B(\pi)\right]
C_{(D-1)} \cr
+B^1(\pi)B^2(\pi)\left[C^1_{(D-1)}, C^2_{(D-1)}\right].
\end{eqnarray}
The first two terms can not contain of ordering ambiguoutes. So the
only source of the anomaly can be commutators of $(D-1)$-dimensional
generators (\ref{16}). The dangerous commutator is known to be
$\left[ {\bf J}^{i-}_{(D-1)}, {\bf J}^{j-}_{(D-1}\right]$, which must
be zero. Its manifest form is
\begin{eqnarray}\label{18}
[{\bf J}^{i-}_{(D-1)},~{\bf J}^{j-}_{(D-1)}]=
\frac{1}{(p^+)^2}
\left[(L_0^{tr}-\bar L_0^{tr}+a)S^{ij}-
(L_0^{tr}-\bar L_0^{tr}-a)\bar S^{ij}\right.+ \cr
\left.\sum^{\infty}_{n=1}[
\frac{D-3}{12}(n-\frac{1}{n})-2n](\alpha^{\left[ i\right.}_{-n}
\alpha^{\left. j\right]}_n+
\bar\alpha^{\left[ i\right.}_{-n}\bar\alpha^{\left. j\right]}_n)\right],
\end{eqnarray}
which is actually zero on the constraint surface (\ref{13}) and under
the conditions
\begin{eqnarray}\label{19}
D=27, \qquad a=2. 
\end{eqnarray}
Note that in terms of the variables (\ref{7}) the same critical 
dimension arises immediately in the old covariant quantization 
framework also, since the no-ghost theorem can be applied without 
modifications to Eqs.(\ref{12}), (\ref{13}).

Let us discuss action which reproduces the Hamiltonian system
(\ref{1})-(\ref{3}). It is convenient to start from the formulation
in terms of the variables (\ref{7}). Then the theory is specified by
$(D-1)$-dimensional Virasoro constraints (\ref{12}), (\ref{13}) for
$x^\mu$ and by the constraint $\pi^N\pi^N=const$ for the additional
vector variable. It prompts to consider action of $(D-1)$-dimensional
string with multiplet of $D$ $\Theta$-terms added 
\footnote{Note that string with one $\Theta$-term added is known to be
equivalent to $D$-string (see [15,16] and references therein),
where it can be easily taken into
account in the path integral framework. It can be clue to
understanding of its appearance in the theory (\ref{20}).}.
\begin{eqnarray}\label{20}
S=S_{(D-1)}-n^N\epsilon^{ab}\partial_aA^N_b-\frac{1}{\phi}
(n^2+\varepsilon).
\end{eqnarray}
Note that the last term can be in fact omitted, since the only which 
is really necessary for the present construction is the condition 
$n^2\ne 0$.
$U(1)^D$ gauge invariance can be used to remove all modes of
$A^N_a, ~ n^N$ except the zero one: $A^N_0=0, ~
A^N_1(\tau, \sigma)=y^N+\pi^N\tau, ~ n^N(\tau, \sigma)=\pi^N$.
While the action has only manifest $(D-1)$ Poincare invariance,
Eq.(\ref{15}) shows that it has also hidden $D$-dimensional
Poincare symmetry. So one expects that it can be rewritten in a
manifestly $D$-dimensional Poincare invariant form. The relevant
action is
\begin{eqnarray}\label{21}
S=\frac{1}{4\pi}\int d^2\sigma \left[\frac{-g^{ab}}{2\sqrt{-g}}
D_ax^N D_bx^N-n^N\epsilon^{ab}\partial_aA^N_b-
\frac{1}{\phi}(n^2+\varepsilon)\right],
\end{eqnarray}
where $D_ax^N\equiv\partial_ax^N-\xi_an^N$. The local symmetries are
$d=2$ reparametrizations, Weyl symmetry and the following
transformations with the parameters $\gamma, ~ \alpha^N, ~ \omega_a$
\begin{eqnarray}\label{22}
\delta x^N=\gamma n^N, \qquad \delta\xi_a=\partial_a\gamma, \qquad
\delta A^N_a=\gamma\frac{\epsilon_{ab}g^{bc}}{\sqrt{-g}}D_cx^N;
\end{eqnarray}
\begin{eqnarray}\label{23}
\delta A^N_a=\partial_a\alpha^N+\omega_an^N, \qquad \delta\phi=
\frac 12\phi^2\epsilon^{ab}\partial_a\omega_b.
\end{eqnarray}
Hamiltonian analysis for the theory is similar to the one presented
in [6]. After partial fixation of gauge, the theory can be formulated
in terms of the phase space variables $x^N(\tau, \sigma), ~
p^N(\tau, \sigma), ~ y^N, ~ \pi^N$ which are subject to the first class
constraints
\begin{eqnarray}\label{24}
\left(p^N\pm\frac{1}{4\pi}\Pi^N{}_M\partial_1x^M\right)^2=0, \cr
\pi^N\pi^N+\varepsilon=0, \qquad \pi^Np^N=0.
\end{eqnarray}
An appropriate gauge for the last constraint turns out to be 
\begin{eqnarray}\label{25}
\pi^Nx^N=0.
\end{eqnarray}
The equations (\ref{24}), (\ref{25}) are equivalent
to Eqs.(\ref{1})-(\ref{3}).

To conclude, it was demonstrated that the light-cone quantization of the 
theory (\ref{21}) is possible, in particular, 
requirement of absence of anomaly in the light-cone Poincare algebra 
leads to the critical dimension $D=27$. There is analogy between the 
action (\ref{21}) and $D$-string which can be clue for understanding 
of $n^N$-dependent part of spectrum. Let us note also that analysis of 
spectrum in the light-cone gauge is more complicated as compare with the 
standard case. In the gauge considered the manifest symmetry is $SO(D-3)$ 
while the massive states should fall into representations of the little 
group $SO(D-1)$. Similar situation arise for $D=11$ membrane 
[17,18] and was analyzed in [8]. It was demonstrated that $SO(8)$ 
multiplets of the first massive level for the toroidal supermembrane 
fall actually into representations of $SO(10)$ group. We hope that the 
analogous consideration is applicable to D=11 superstring also.

\section*{Acknowledgments.}

The work has been supported by FAPERJ and partially by
Project INTAS-96-0308 and by Project GRACENAS 
No 97-6.2-34.

\end{document}